\documentclass[showpacs,oneside,twocolumn,prl,amsmath,amssymb, fleqn]{revtex4-1}
\usepackage{cases}
\usepackage{amsmath}
\usepackage{amssymb}
\usepackage{amsfonts}
\usepackage{amssymb}
\usepackage{dcolumn}
\usepackage{bm}
\usepackage{bbm}
\usepackage{hyperref}
\hypersetup{colorlinks=true,breaklinks=true,pdfstartview=FitH,linkcolor=blue,citecolor=green}
\usepackage{graphicx}
\usepackage{xcolor}
\usepackage{array}
\usepackage{subfigure}
\usepackage{ulem}

\begin{document}
\title{Quantum Discord for Investigating Quantum Correlations without Entanglement in Solids}
\author{Xing Rong}
\author{Zixiang Wang}
\author{Fangzhou Jin}
\author{Jianpei Geng}
\author{Pengbo Feng}
\author{Nanyang Xu}
\author{Ya Wang}
\author{Chenyong Ju}
\author{Mingjun Shi}
\altaffiliation{shmj@ustc.edu.cn}
\author{Jiangfeng Du}
\altaffiliation{djf@ustc.edu.cn}
\affiliation{Hefei National Laboratory for Physics Sciences at Microscale and Department of Modern Physics, University of Science
and Technology of China, Hefei, 230026, China}

\begin{abstract}
Quantum systems unfold diversified correlations which have no classical counterparts. These quantum correlations have various different facets.
Quantum entanglement, as the most well known measure of quantum correlations, plays essential roles in quantum information processing. However, it has recently been pointed out that quantum entanglement cannot describe all the nonclassicality in the correlations.
Thus the study of quantum correlations in separable states attracts widely attentions.
Herein, we experimentally investigate the quantum correlations of separable thermal states in terms of quantum discord.
The sudden change of quantum discord is observed, which captures ambiguously the critical point associated with the behavior of Hamiltonian. Our results display the potential applications of quantum correlations in studying the fundamental properties of quantum system, such as quantum criticality of non-zero temperature.
\end{abstract}

\maketitle

Nowadays the significance of quantum correlations goes well beyond quantum entanglement, which has been widely investigated and believed to be the key resource of quantum information processing \cite{RevModPhys.81.865}. However, quantum correlations possess other facets for which the quantum entanglement does not provide a complete characterization: there exist quantum nonlocality without entanglement \cite{Bennett.PhysRevA.59.1070.1999}.
Moreover, quantum entanglement is not a necessary for quantum computation: we presented the experiment to perform a quantum algorithm for parity problem  without using entanglement as early as 2001 \cite{PhysRevA.64.042306}, and recently a deterministic quantum computation with one qubit (DQC1 \cite{PhysRevLett.81.5672}) is realized experimentally \cite{PhysRevLett.101.200501} on the mixed separable states. 
In this context, it is actually the type of quantum correlation known as quantum discord
\cite{Ollivier:2002ly,Henderson.JPhysA.34.6899.2001}, rather than quantum entanglement, that provides the enhancement for the computation
\cite{PhysRevLett.100.050502}.
The significance of quantum discord in quantum communication has been studied in the cases of local broadcasting \cite{Piani.PhysRevLett.100.090502.2008} and state merging
\cite{Cavalcanti.PhysRevA.83.032324.2011,*Madhok.PhysRevA.83.032323.2011}.

Quantum correlations also provide a powerful framework for the understanding of complex quantum systems.
The first step towards this investigation is due to Osborne and Nielsen
\cite{Osborne.PhysRevA.66.032110}, who studied the relation between the entanglement and the quantum phase transition in $XY$ model.
Recently, it is shown that quantum discord, in contrast to entanglement, captures the critical points associated with quantum phase transitions for $XXZ$ and $XY$ model, even at finite temperature or in an external magnetic field
\cite{Dillenschneider:2008zr,PhysRevA.80.022108,PhysRevLett.105.095702,PhysRevA.82.012106,PhysRevA.83.062334}.
Besides, quantum correlations are of great importance in quantum thermodynamics \cite{PhysRevA.67.012320,RevModPhys.81.1}, and even in photosynthesis \cite{Sarovar:2010fk}.
Thus the study of the quantum correlations will be of both fundamental and practical significance.

In this Letter, we report an experiment to characterize the quantum correlations in separable thermal states.
Quantum discord is used as the quantifier of the quantumness in the correlation in solids.
A series of separable thermal states are generated,
which can be taken as the thermal-equilibrium states of a two-qubit $XXZ$ Heisenberg model at a finite temperature.
By tuning the state parameter that corresponds to the anisotropic coupling constant in the $XXZ$ model,
we observe the sudden changes of quantum discord.
The sudden change corresponds exactly to the energy-level crossing of the $XXZ$ Hamiltonian,
and also indicates the critical point on which the ground state of the Hamiltonian transforms from product state to entangled one, or vice versa.
When the number of the qubit on the $XXZ$ chain tends to infinity (i.e., thermodynamic limit),
the sudden changes of discord spotlight the critical points associated with quantum phase transitions \cite{PhysRevLett.105.095702}.
Thus our experiment opens the possibility of experimentally studying the fundamental properties of quantum system from the viewpoint of quantum correlation, in particular quantum discord.

The thermal states generated in our experiment take the Bell-diagonal form, that is,
\begin{equation}\label{Bell-diagonal-states}
  \rho_{\mathrm{BD}}= \frac{1}{4}\big(\mathbbm{1}
  +c_x\sigma_x^1\sigma_x^2+c_y\sigma_y^1\sigma_y^2+c_z\sigma_z^1\sigma_z^2\big),
\end{equation}
where $\sigma_i^{1(2)}$ ($i=x,y,z$) are the Pauli matrices of the first (second) qubit.
From \cite{PhysRevA.77.042303}, the quantum discord of $\rho_{\mathrm{BD}}$ is given by
\begin{equation}\label{discord}
  D(\rho_{\mathrm{BD}})=1+h(c)-S(\rho_{\mathrm{BD}}),
\end{equation}
where $c=\max\{|c_x|,|c_y|,|c_z|\}$, $S(\rho)=-\mathrm{Tr}(\rho\log_2\rho)$ is the von Neumann entropy,
and the function $h(x)$ is defined as
$h(x)=-\frac{1+x}{2}\log_2\frac{1+x}{2}-\frac{1-x}{2}\log_2\frac{1-x}{2}$.

\begin{figure}[htbp]
	\centering\includegraphics[width=1\columnwidth]{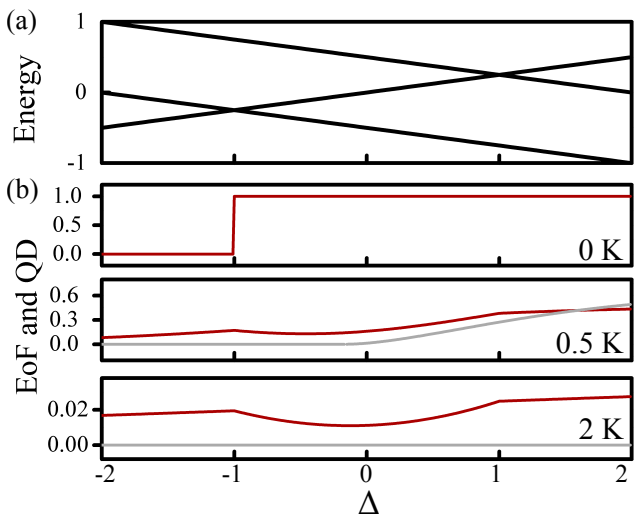}
	\caption{(color online) (a) the energy-level of the Hamiltonian given by Eq. \eqref{Hamiltonian}.
(b) the EoF (grey lines) and quantum discord (red lines) of the state $\rho$ given by Eq. \eqref{thermal states}.
When the temperature increase from $0$K to $2$K, the EoF vanishes and the discord remains positive.
The sudden changes of discord correspond to the level crossing shown in (a).}
	\label{fig1}
\end{figure}

The Bell-diagonal states of the form \eqref{Bell-diagonal-states} can be taken as the thermal-equilibrium states of a two-qubit $XXZ$ Heisenberg chain with the Hamiltonian given by
\begin{equation}\label{Hamiltonian}
H =\frac{J}{4}(\sigma_x^1\sigma_x^2+\sigma_y^1\sigma_y^2+\Delta\sigma_z^1\sigma_z^2),
\end{equation}
where $J$ is the coupling constant and $\Delta$ is anisotropy parameter.
In fact, the density matrix for this model at thermal equilibrium at temperature $T$ is given by the canonical ensemble $\rho=e^{-\beta H}/Z$, where $\beta=1/k_BT$, and $Z=\mathrm{Tr}(e^{-\beta H})$ is the partition function.
When $T=0$, the state $\rho$ is the ground state of the Hamiltonian $H$.
When the temperature $T$ is appreciably larger than the maximum splitting of $H$ in Eq. \eqref{Hamiltonian}, the thermal-equilibrium state can be approximated as
\begin{equation}\label{thermal states}
  \rho=\frac{\mathbbm1}{4}-\frac{\beta J}{16}(\sigma_x^1\sigma_x^2+\sigma_y^1\sigma_y^2+\Delta\sigma_z^1\sigma_z^2).
\end{equation}
The entanglement of formation (EoF) \cite{PhysRevA.53.2046} and the quantum discord of the state $\rho$ can be easily obtained from Ref. \cite{Wootters:1998fk} and Eq. \eqref{discord} respectively.
We depict the results with respect to the anisotropic parameter $\Delta$ in Fig. \ref{fig1},
where the units have been chosen such that $J = 1$ and $k_B$ is unity \cite{PhysRevA.71.012307,*PhysRevLett.105.240405}.
It is shown that with the temperature increasing, the EoF becomes zero, while the quantum discord always remains positive [Fig. \ref{fig1}(b)].
Also, at finite temperature, the quantum discord changes suddenly at $\Delta=\pm1$ which corresponds exactly to the energy-level crossing, but it is not the case for EoF:
The behavior of EoF is illustrated by smooth curves (grey lines in Fig. \ref{fig1}(b)).

\begin{figure}[htbp]
	\centering\includegraphics[width=0.95\columnwidth]{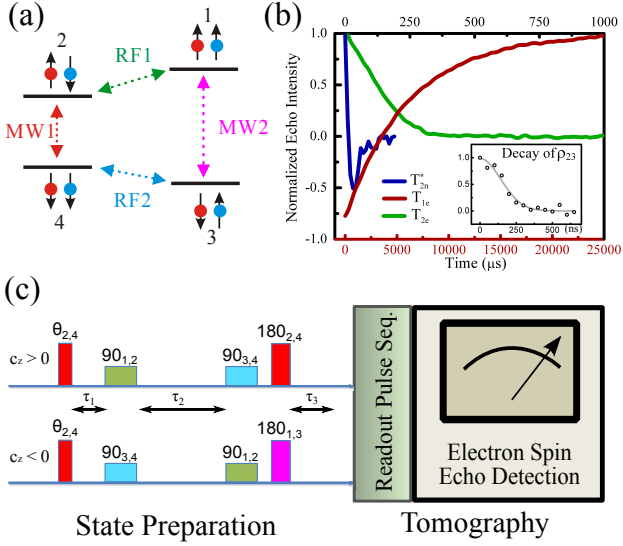}
	\caption{(color online)
			(a) Energy level diagram of the P:Si system. There are four Zeeman product states which are labeled states $1- 4$, where the left and the right $\uparrow$ ($\downarrow$) stands for the $\pm 1/2$ states of electron and nuclear spins, respectively. EPR and NMR transitions are indicated by two-way arrows.
			(b) Experimental results of the relaxation time measurements. The red line are collected from the echo-detected inversion recovery experiment (the related time axis is the red bottom one), which reveals $T_{1e}$. The green line is the electron spin echo decay through the Hahn echo sequence, which gives $T_{2e}$. The blue curve is the result of the nuclear spin FID with pulse sequence ($\pi_{2,4} - \pi/2_{1,2}-\tau-\pi/2_{1,2}-$echo, $\tau$ a variable), which is used to calculate $T_{2n}^*$. Note that the time axis for the green and blue curves is the black top one. The inset shows the decay of $\rho_{23}$. The experimental data (black circles) were fitted with a damped exponential function (grey line) of the form $y = y_0 + Ae^{-(\frac{t}{t_c})^2}$.
			(c) Diagram of the experimental pulse sequence. We use upper (lower) pulse sequence to generate the state $\rho$, in the case of $c_z > 0$ ($c_z < 0$). $\theta_{2,4}$ ($180_{1,3}$) is a MW$1$(MW$2$) pulse to rotate the electron spin by $\theta$ ($\pi$). $90_{1,2(3,4)}$ stands for a $\pi/2$ RF$1$(RF$2$) pulse. The tomography procession is composed of readout pulse sequence and electron spin echo detection.
			}
	\label{fig2}
\end{figure}

To explore the quantum correlations in the separable states, we choose phosphorous donors in silicon (P:Si) material \cite{PhysRevLett.106.040501,PhysRevB.68.193207} with P concentration about $1 \times 10^{16}$cm$^{-3}$ as a benchmark system. The system consists of an electron spin $S = 1/2$ and a nuclear spin $I = 1/2$. It can be described by an isotropic spin Hamiltonian:
 \begin{equation}
 \label{H of P:Si}
  H_{e,n} = \omega_e S_z - \omega_I I_z +2\pi a\cdot \overrightarrow{\textbf{S}}\cdot\overrightarrow{\textbf{I}},
  \end{equation}
where $\omega_e = g\beta_e B_0/\hbar$ and $\omega_I = g_I\beta_IB_0/\hbar$ characterize the Zeeman interaction for the electron and nuclear spins, $a=117~MHz$ is the isotropic hyperfine interaction strength and $\overrightarrow{\textbf{S}}$($\overrightarrow{\textbf{I}}$) is the electron (nuclear) spin operator. The energy diagram of this system is plotted in Fig. \ref{fig2}(a), where the two electron-paramagnetic-resonance (EPR) transitions and two NMR transitions have been labeled by MW1, MW2 and RF1 and RF2 respectively. The frequencies of these transitions, which are determined by field-swept echo detection (FSED) and standard Davies electron-nuclear double resonance (ENDOR) experiments, are MW$1 = 9.701~$GHz, MW$2=9.818~$GHz, RF$1=52.383~$MHz and RF$2=65.181~$MHz. The pulse length of the EPR (NMR) $\pi$ pulse is $60~$ns ($10~\mu$s) which has been derived from the EPR (NMR) Rabi nutation experiment.

The experiment has been carried out at $T_{\rm{exp}} = 8~$K on a homebuilt pulsed electron spin resonance (ESR) spectrometer, which provides the access to controlling both the electron and nuclear spins with flexible microwave (MW) and radio frequency (RF) pulses\cite{PhysRevA.63.042302}. 
We first measure the relaxation times of both electron and nuclear spins (see Fig. \ref{fig2}(b)). 
For the electron spin, the transverse relaxation time is $T_{2e} = 120~\mu$s (green line in Fig. \ref{fig2}(b)) and the electron population relaxation time is $T_{1e} = 5.6~$ms (red line in Fig. \ref{fig2}(b)). 
For the nuclear spin, the dephasing time is determined as $T_{2n}^* = 24.3~\mu s$ by a nuclear spin free induction decay (FID) experiment (blue line in Fig. \ref{fig2}(b)). We also measure the decay of the $\rho_{23}$ under the environment with a decay constant $t_c = 200~$ns (plotted in the inset of Fig. \ref{fig2}(b)). 
The relaxation times are similar to those reported in Ref. \cite{PhysRevLett.106.040501}. 
The nuclear population relaxation time $T_{1n}$ is estimated to be 250 times of $T_{1e}$ according to the literatures \cite{Tyryshkin:2006fk}. So the waiting time between each experiment has been set to $10~$s.

Fig. \ref{fig2}(c) shows the pulse sequence applied in our experiment. 
The first MW pulse is to flip the electron spin with an angle $\theta$ while $I_z = -1/2$. 
There is a waiting time ($\tau_1 = 1~\mu$s $\gg T_{2e}^*$) following so that the off-diagonal elements of the density matrix decay off. 
The first $90^\circ$ RF1(RF2) pulse is to equalize the diagonal elements $\rho_{11}$ and $\rho_{22}$ ($\rho_{33}$ and $\rho_{44}$). 
Then we let the off-diagonal elements, which are generated by the first RF pulse, decay off by another waiting time $\tau_2 = 200~\mu$s which is long enough compared with $T_{2n}^* $. 
The following $90^\circ$ RF2(RF1) pulse ($90^\circ$) and the $180^\circ$ MW1(MW2) pulses are to generate the nonzero off-diagonal element, $\rho_{23}$. 
After waiting time $\tau_3$ which is to introduce a decay factor $\lambda(\tau_3)$, the final state has been prepared to Bell diagonal states given by Eq. \eqref{Bell-diagonal-states}, where $c_x = c_y = -\epsilon(1-\cos\theta)\lambda(\tau_3)$, $c_z = \pm 2\epsilon(1+\cos\theta)$ and $\epsilon = g\beta_e B_0/ 8 k_B T_{\rm{exp}} = 7.35\times 10^{-3}$. Here $\lambda(\tau_3) = \exp(-\tau_3/t_c)$ which has been determined in Fig. \ref{fig2}(b).

 \begin{figure}[htbp]
	\centering
			 \includegraphics[width=1\columnwidth]{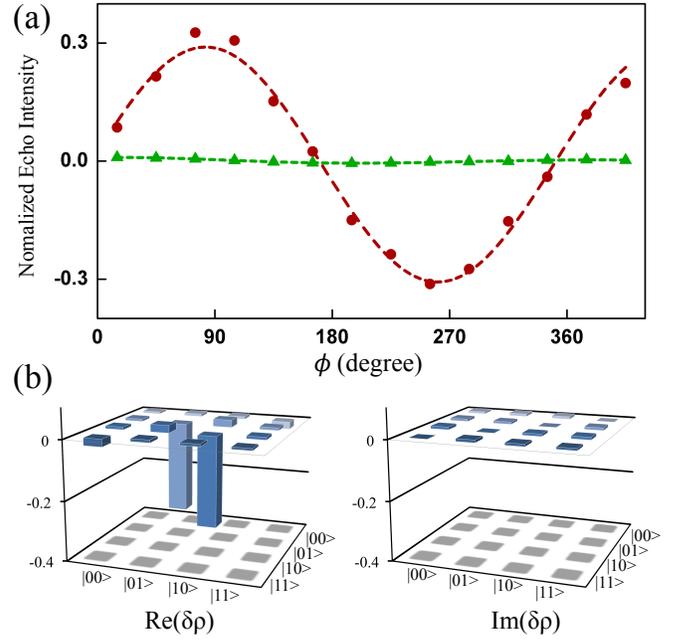}
	\caption{(color online) (a) The readout of the $\rho_{23}$. I$_{+x} -$I$_{-x}$ (red circles) and I$_{+y} -$I$_{-y}$(green triangles) are plotted. Sine functions(dashed lines) are fitted to the experimental data. The x axis is the rotation angle $\phi$ and the y axis is the amplitude of the echo in unit of $\epsilon$. The results give both real and imaginary parts of $\rho_{23}$.
			(b)The reconstructed deviation density matrix $\delta \rho = \rho - \frac{1}{4} \mathbbm{1}$ in unit of $\epsilon$, when $c_x = c_y = -0.0044$ and $c_z = 0.0008$.
			}
	\label{fig3}
\end{figure}

Tomography technique \cite{Vandersypen:2004fk} is then used to reconstruct the density matrix $\rho$. 
By observing both the in-phase and quadrature components of the electron spin echo (the block labeled by electron spin echo detection in Fig. \ref{fig2}(c)), measurement in $S_x$ and $S_y$ bases is achieved. 
For the diagonal elements $\rho_{ii, ~i = 1-4}$, the readout pulse sequence is a MW$1$(2) (RF$1$(2)) pulse (green block in Fig. \ref{fig2}(c)) with variable duration. 
Note that an additional waiting time $\tau_4 = 200~\mu$s (not shown in Fig. \ref{fig2}(c)), which is long enough to let dephasing effect eliminate the possible off-diagonal elements, is inserted before the readout pulse. 
Then this selective electron(nuclear) spin Rabi nutation corresponds to the measurement of $S_zI^{\alpha, \beta}$($S^{\alpha, \beta}I_z$) where $S(I)^{\alpha, \beta} = (1 \pm \sigma_z)/2$. To readout the off-diagonal elements $\rho_{12}$, $\rho_{34}$, $\rho_{13}$ and $\rho_{24}$, $\tau_{4}$ is set to zero. Compared with the previous nutation experimental data, these off-diagonal elements were calculated to be almost zero in our experiment.

To measure $\rho_{23}$, the readout pulse sequence is composed of a microwave pulse, $180_{2,4}$, and a radio frequency pulse, $\phi_{3,4}$, with variable rotation angle $\phi$. We denote the spectrum of Rabi nutation, when the phase of $180_{2,4}$ is $\pm x, \pm y$, as $I_{\pm x, \pm y}(\phi)$. After some calculations, we find that $I_{+x}(\phi) - I_{-x}(\phi)=- Re(\rho_{23})\sin(\phi)$ and $I_{+y}(\phi) - I_{-y}(\phi)= Im(\rho_{23})\sin(\phi)$ [Fig. \ref{fig3}(a)]. For the measurement of $\rho_{14}$, we use $\phi_{1,2}$ instead of $\phi_{3,4}$ in the readout pulse sequence of $\rho_{23}$. To normalize the results, we compare all the the amplitudes of the Rabi nutations in our tomography procession to the amplitude of electron spin Rabi nutation which can be taken as $2\epsilon$. Once the above steps are completed, the experimental density matrix can be fully reconstructed. Fig. \ref{fig3}(b) shows the result of the tomography of one Bell diagonal state whose $c_x = c_y = -0.0044$ and $c_z = 0.0008$ (Note that $\delta\rho$ is the result the reconstructed density matrix subtracted by $\mathbbm{1}/4$). The entanglement of this quantum state is zero, thus the state is separable. However, the quantum discord is $1.45\times 10^{-5}$. This shows that nonzero quantum correlations have been generated in a separable quantum state.
 \begin{figure}[htbp]
	\centering
			 \includegraphics[width=1\columnwidth]{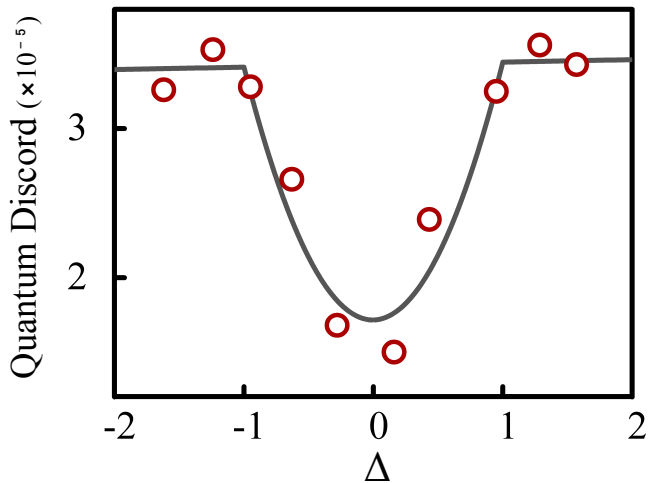}
	\caption{(color online) The behavior of quantum correlations when $\Delta$ varies. The $x$ axis is the tuning parameter $\Delta$ and the $y$ axis is quantum discord. Experimental data of quantum discord (red circles), which are derived from the reconstructed density matrices, agree with the theoretical prediction using Eq. \eqref{discord} (solid line). The deviation between the experimental data and the theoretical prediction is due to the imperfection of the pulses.
	}
	\label{fig4}
\end{figure}

The values of quantum discord (red circles in Fig. \ref{fig4}(b)) are numerically calculated using Eq. \eqref{discord}.
The experimental data agree with the theoretical prediction (line in the Fig. \ref{fig4}(b)). It is clear that quantum discord experiences a sudden change at $\Delta=\pm1$ where the energy level crossing occurs. It is worthy of pointing out that the entanglement, which is characterized by EoF, is measured to be zero in our experiments. Thus, we have presented a case where entanglement fails to capture energy level crossing while temperature is of a finite value. Note that there are theoretical works which conclude that quantum discord can be utilized to capture the quantum phase transition even at finite temperatures\cite{PhysRevLett.105.095702,PhysRevA.83.062334,PhysRevA.81.044101}. The success in highlighting the sudden change of the ground state ($\Delta = -1$) in the two-qubit system employed in our experiments indicates that quantum correlation could be used to observe the quantum criticality at finite temperatures.

In summary, we have experimentally observed quantum correlations in a series of separable states in solids. Furthermore, the capability of using quantum correlations to reveal the intrinsic change of the physical system is revealed. Two-qubit XXZ Heisenberg model has been taken as an example, and the abrupt change of its ground state has been unambiguously spotlighted by quantum correlations. Our experiment may serve as a preliminary meaningful step to observe quantum criticality at finite temperatures via quantum correlations.

We acknowledge X. H. Peng, C. J. Zhang and C. K. Duan for helpful discussion. This work was supported by National Nature Science Foundation of China (Grants Nos.10834005, 91021005, and 21073171), the Instrument Developing Project of the Chinese Academy of Sciences (Grant No. Y2010025), and the National Fundamental Research Program 2007CB925200.

\bibliography{myref}

\begin{thebibliography}{31}%
\makeatletter
\providecommand \@ifxundefined [1]{%
 \@ifx{#1\undefined}
}%
\providecommand \@ifnum [1]{%
 \ifnum #1\expandafter \@firstoftwo
 \else \expandafter \@secondoftwo
 \fi
}%
\providecommand \@ifx [1]{%
 \ifx #1\expandafter \@firstoftwo
 \else \expandafter \@secondoftwo
 \fi
}%
\providecommand \natexlab [1]{#1}%
\providecommand \enquote  [1]{``#1''}%
\providecommand \bibnamefont  [1]{#1}%
\providecommand \bibfnamefont [1]{#1}%
\providecommand \citenamefont [1]{#1}%
\providecommand \href@noop [0]{\@secondoftwo}%
\providecommand \href [0]{\begingroup \@sanitize@url \@href}%
\providecommand \@href[1]{\@@startlink{#1}\@@href}%
\providecommand \@@href[1]{\endgroup#1\@@endlink}%
\providecommand \@sanitize@url [0]{\catcode `\\12\catcode `\$12\catcode
  `\&12\catcode `\#12\catcode `\^12\catcode `\_12\catcode `\%12\relax}%
\providecommand \@@startlink[1]{}%
\providecommand \@@endlink[0]{}%
\providecommand \url  [0]{\begingroup\@sanitize@url \@url }%
\providecommand \@url [1]{\endgroup\@href {#1}{\urlprefix }}%
\providecommand \urlprefix  [0]{URL }%
\providecommand \Eprint [0]{\href }%
\@ifxundefined \urlstyle {%
  \providecommand \doi  [0]{\begingroup \@sanitize@url \@doi}%
  \providecommand \@doi [1]{\endgroup \@@startlink {\doibase
  #1}doi:\discretionary {}{}{}#1\@@endlink }%
}{%
  \providecommand \doi  [0]{doi:\discretionary{}{}{}\begingroup
  \urlstyle{rm}\Url }%
}%
\providecommand \doibase [0]{http://dx.doi.org/}%
\providecommand \Doi [0]{\begingroup \@sanitize@url \@Doi }%
\providecommand \@Doi  [1]{\endgroup\@@startlink{\doibase#1}\@@Doi}%
\providecommand \@@Doi [1]{#1\@@endlink}%
\providecommand \selectlanguage [0]{\@gobble}%
\providecommand \bibinfo  [0]{\@secondoftwo}%
\providecommand \bibfield  [0]{\@secondoftwo}%
\providecommand \translation [1]{[#1]}%
\providecommand \BibitemOpen [0]{}%
\providecommand \bibitemStop [0]{}%
\providecommand \bibitemNoStop [0]{.\EOS\space}%
\providecommand \EOS [0]{\spacefactor3000\relax}%
\providecommand \BibitemShut  [1]{\csname bibitem#1\endcsname}%
\bibitem [{\citenamefont {Horodecki}\ \emph {et~al.}(2009)\citenamefont
  {Horodecki}, \citenamefont {Horodecki}, \citenamefont {Horodecki},\ and\
  \citenamefont {Horodecki}}]{RevModPhys.81.865}%
  \BibitemOpen
  \bibfield  {author} {\bibinfo {author} {\bibfnamefont {R.}~\bibnamefont
  {Horodecki}}, \bibinfo {author} {\bibfnamefont {P.}~\bibnamefont
  {Horodecki}}, \bibinfo {author} {\bibfnamefont {M.}~\bibnamefont
  {Horodecki}}, \ and\ \bibinfo {author} {\bibfnamefont {K.}~\bibnamefont
  {Horodecki}},\ }\Doi {10.1103/RevModPhys.81.865} {\bibfield  {journal}
  {\bibinfo  {journal} {Rev. Mod. Phys.},\ }\textbf {\bibinfo {volume} {81}},\
  \bibinfo {pages} {865} (\bibinfo {year} {2009})}\BibitemShut {NoStop}%
\bibitem [{\citenamefont {Bennett}\ \emph {et~al.}(1999)\citenamefont
  {Bennett}, \citenamefont {DiVincenzo}, \citenamefont {Fuchs}, \citenamefont
  {Mor}, \citenamefont {Rains}, \citenamefont {Shor}, \citenamefont {Smolin},\
  and\ \citenamefont {Wootters}}]{Bennett.PhysRevA.59.1070.1999}%
  \BibitemOpen
  \bibfield  {author} {\bibinfo {author} {\bibfnamefont {C.~H.}\ \bibnamefont
  {Bennett}}, \bibinfo {author} {\bibfnamefont {D.~P.}\ \bibnamefont
  {DiVincenzo}}, \bibinfo {author} {\bibfnamefont {C.~A.}\ \bibnamefont
  {Fuchs}}, \bibinfo {author} {\bibfnamefont {T.}~\bibnamefont {Mor}}, \bibinfo
  {author} {\bibfnamefont {E.}~\bibnamefont {Rains}}, \bibinfo {author}
  {\bibfnamefont {P.~W.}\ \bibnamefont {Shor}}, \bibinfo {author}
  {\bibfnamefont {J.~A.}\ \bibnamefont {Smolin}}, \ and\ \bibinfo {author}
  {\bibfnamefont {W.~K.}\ \bibnamefont {Wootters}},\ }\Doi
  {10.1103/PhysRevA.59.1070} {\bibfield  {journal} {\bibinfo  {journal} {Phys.
  Rev. A},\ }\textbf {\bibinfo {volume} {59}},\ \bibinfo {pages} {1070}
  (\bibinfo {year} {1999})}\BibitemShut {NoStop}%
\bibitem [{\citenamefont {Du}\ \emph {et~al.}(2001){\natexlab{a}}\citenamefont
  {Du}, \citenamefont {Shi}, \citenamefont {Zhou}, \citenamefont {Fan},
  \citenamefont {Ye}, \citenamefont {Han},\ and\ \citenamefont
  {Wu}}]{PhysRevA.64.042306}%
  \BibitemOpen
  \bibfield  {author} {\bibinfo {author} {\bibfnamefont {J.}~\bibnamefont
  {Du}}, \bibinfo {author} {\bibfnamefont {M.}~\bibnamefont {Shi}}, \bibinfo
  {author} {\bibfnamefont {X.}~\bibnamefont {Zhou}}, \bibinfo {author}
  {\bibfnamefont {Y.}~\bibnamefont {Fan}}, \bibinfo {author} {\bibfnamefont
  {B.}~\bibnamefont {Ye}}, \bibinfo {author} {\bibfnamefont {R.}~\bibnamefont
  {Han}}, \ and\ \bibinfo {author} {\bibfnamefont {J.}~\bibnamefont {Wu}},\
  }\Doi {10.1103/PhysRevA.64.042306} {\bibfield  {journal} {\bibinfo  {journal}
  {Phys. Rev. A},\ }\textbf {\bibinfo {volume} {64}},\ \bibinfo {pages}
  {042306} (\bibinfo {year} {2001}{\natexlab{a}})}\BibitemShut {NoStop}%
\bibitem [{\citenamefont {Knill}\ and\ \citenamefont
  {Laflamme}(1998)}]{PhysRevLett.81.5672}%
  \BibitemOpen
  \bibfield  {author} {\bibinfo {author} {\bibfnamefont {E.}~\bibnamefont
  {Knill}}\ and\ \bibinfo {author} {\bibfnamefont {R.}~\bibnamefont
  {Laflamme}},\ }\Doi {10.1103/PhysRevLett.81.5672} {\bibfield  {journal}
  {\bibinfo  {journal} {Phys. Rev. Lett.},\ }\textbf {\bibinfo {volume} {81}},\
  \bibinfo {pages} {5672} (\bibinfo {year} {1998})}\BibitemShut {NoStop}%
\bibitem [{\citenamefont {Lanyon}\ \emph {et~al.}(2008)\citenamefont {Lanyon},
  \citenamefont {Barbieri}, \citenamefont {Almeida},\ and\ \citenamefont
  {White}}]{PhysRevLett.101.200501}%
  \BibitemOpen
  \bibfield  {author} {\bibinfo {author} {\bibfnamefont {B.~P.}\ \bibnamefont
  {Lanyon}}, \bibinfo {author} {\bibfnamefont {M.}~\bibnamefont {Barbieri}},
  \bibinfo {author} {\bibfnamefont {M.~P.}\ \bibnamefont {Almeida}}, \ and\
  \bibinfo {author} {\bibfnamefont {A.~G.}\ \bibnamefont {White}},\ }\Doi
  {10.1103/PhysRevLett.101.200501} {\bibfield  {journal} {\bibinfo  {journal}
  {Phys. Rev. Lett.},\ }\textbf {\bibinfo {volume} {101}},\ \bibinfo {pages}
  {200501} (\bibinfo {year} {2008})}\BibitemShut {NoStop}%
\bibitem [{\citenamefont {Ollivier}\ and\ \citenamefont
  {Zurek}(2002)}]{Ollivier:2002ly}%
  \BibitemOpen
  \bibfield  {author} {\bibinfo {author} {\bibfnamefont {H.}~\bibnamefont
  {Ollivier}}\ and\ \bibinfo {author} {\bibfnamefont {W.}~\bibnamefont
  {Zurek}},\ }\Doi {DOI 10.1103/PhysRevLett.88.017901} {\bibfield  {journal}
  {\bibinfo  {journal} {Physical Review Letters},\ }\textbf {\bibinfo {volume}
  {88}},\ \bibinfo {pages} {017901} (\bibinfo {year} {2002})}\BibitemShut
  {NoStop}%
\bibitem [{\citenamefont {Henderson}\ and\ \citenamefont
  {Vedral}(2001)}]{Henderson.JPhysA.34.6899.2001}%
  \BibitemOpen
  \bibfield  {author} {\bibinfo {author} {\bibfnamefont {L.}~\bibnamefont
  {Henderson}}\ and\ \bibinfo {author} {\bibfnamefont {V.}~\bibnamefont
  {Vedral}},\ }\Doi {10.1088/0305-4470/34/35/315} {\bibfield  {journal}
  {\bibinfo  {journal} {J. Phys. A: Math. Gen.},\ }\textbf {\bibinfo {volume}
  {34}},\ \bibinfo {pages} {6899} (\bibinfo {year} {2001})}\BibitemShut
  {NoStop}%
\bibitem [{\citenamefont {Datta}\ \emph {et~al.}(2008)\citenamefont {Datta},
  \citenamefont {Shaji},\ and\ \citenamefont {Caves}}]{PhysRevLett.100.050502}%
  \BibitemOpen
  \bibfield  {author} {\bibinfo {author} {\bibfnamefont {A.}~\bibnamefont
  {Datta}}, \bibinfo {author} {\bibfnamefont {A.}~\bibnamefont {Shaji}}, \ and\
  \bibinfo {author} {\bibfnamefont {C.~M.}\ \bibnamefont {Caves}},\ }\Doi
  {10.1103/PhysRevLett.100.050502} {\bibfield  {journal} {\bibinfo  {journal}
  {Phys. Rev. Lett.},\ }\textbf {\bibinfo {volume} {100}},\ \bibinfo {pages}
  {050502} (\bibinfo {year} {2008})}\BibitemShut {NoStop}%
\bibitem [{\citenamefont {Piani}\ \emph {et~al.}(2008)\citenamefont {Piani},
  \citenamefont {Horodecki},\ and\ \citenamefont
  {Horodecki}}]{Piani.PhysRevLett.100.090502.2008}%
  \BibitemOpen
  \bibfield  {author} {\bibinfo {author} {\bibfnamefont {M.}~\bibnamefont
  {Piani}}, \bibinfo {author} {\bibfnamefont {P.}~\bibnamefont {Horodecki}}, \
  and\ \bibinfo {author} {\bibfnamefont {R.}~\bibnamefont {Horodecki}},\ }\Doi
  {10.1103/PhysRevLett.100.090502} {\bibfield  {journal} {\bibinfo  {journal}
  {Phys. Rev. Lett.},\ }\textbf {\bibinfo {volume} {100}},\ \bibinfo {pages}
  {090502} (\bibinfo {year} {2008})}\BibitemShut {NoStop}%
\bibitem [{\citenamefont {Cavalcanti}\ \emph {et~al.}(2011)\citenamefont
  {Cavalcanti}, \citenamefont {Aolita}, \citenamefont {Boixo}, \citenamefont
  {Modi}, \citenamefont {Piani},\ and\ \citenamefont
  {Winter}}]{Cavalcanti.PhysRevA.83.032324.2011}%
  \BibitemOpen
  \bibfield  {author} {\bibinfo {author} {\bibfnamefont {D.}~\bibnamefont
  {Cavalcanti}}, \bibinfo {author} {\bibfnamefont {L.}~\bibnamefont {Aolita}},
  \bibinfo {author} {\bibfnamefont {S.}~\bibnamefont {Boixo}}, \bibinfo
  {author} {\bibfnamefont {K.}~\bibnamefont {Modi}}, \bibinfo {author}
  {\bibfnamefont {M.}~\bibnamefont {Piani}}, \ and\ \bibinfo {author}
  {\bibfnamefont {A.}~\bibnamefont {Winter}},\ }\Doi
  {10.1103/PhysRevA.83.032324} {\bibfield  {journal} {\bibinfo  {journal}
  {Phys. Rev. A},\ }\textbf {\bibinfo {volume} {83}},\ \bibinfo {pages}
  {032324} (\bibinfo {year} {2011})}\BibitemShut {NoStop}%
\bibitem [{\citenamefont {Madhok}\ and\ \citenamefont
  {Datta}(2011)}]{Madhok.PhysRevA.83.032323.2011}%
  \BibitemOpen
  \bibfield  {author} {\bibinfo {author} {\bibfnamefont {V.}~\bibnamefont
  {Madhok}}\ and\ \bibinfo {author} {\bibfnamefont {A.}~\bibnamefont {Datta}},\
  }\Doi {10.1103/PhysRevA.83.032323} {\bibfield  {journal} {\bibinfo  {journal}
  {Phys. Rev. A},\ }\textbf {\bibinfo {volume} {83}},\ \bibinfo {pages}
  {032323} (\bibinfo {year} {2011})}\BibitemShut {NoStop}%
\bibitem [{\citenamefont {Osborne}\ and\ \citenamefont
  {Nielsen}(2002)}]{Osborne.PhysRevA.66.032110}%
  \BibitemOpen
  \bibfield  {author} {\bibinfo {author} {\bibfnamefont {T.~J.}\ \bibnamefont
  {Osborne}}\ and\ \bibinfo {author} {\bibfnamefont {M.~A.}\ \bibnamefont
  {Nielsen}},\ }\Doi {10.1103/PhysRevA.66.032110} {\bibfield  {journal}
  {\bibinfo  {journal} {Phys. Rev. A},\ }\textbf {\bibinfo {volume} {66}},\
  \bibinfo {pages} {032110} (\bibinfo {year} {2002})}\BibitemShut {NoStop}%
\bibitem [{\citenamefont {Dillenschneider}(2008)}]{Dillenschneider:2008zr}%
  \BibitemOpen
  \bibfield  {author} {\bibinfo {author} {\bibfnamefont {R.}~\bibnamefont
  {Dillenschneider}},\ }\Doi {DOI 10.1103/PhysRevB.78.224413} {\bibfield
  {journal} {\bibinfo  {journal} {Physical Review B},\ }\textbf {\bibinfo
  {volume} {78}},\ \bibinfo {pages} {224413} (\bibinfo {year}
  {2008})}\BibitemShut {NoStop}%
\bibitem [{\citenamefont {Sarandy}(2009)}]{PhysRevA.80.022108}%
  \BibitemOpen
  \bibfield  {author} {\bibinfo {author} {\bibfnamefont {M.~S.}\ \bibnamefont
  {Sarandy}},\ }\Doi {10.1103/PhysRevA.80.022108} {\bibfield  {journal}
  {\bibinfo  {journal} {Phys. Rev. A},\ }\textbf {\bibinfo {volume} {80}},\
  \bibinfo {pages} {022108} (\bibinfo {year} {2009})}\BibitemShut {NoStop}%
\bibitem [{\citenamefont {Werlang}\ \emph {et~al.}(2010)\citenamefont
  {Werlang}, \citenamefont {Trippe}, \citenamefont {Ribeiro},\ and\
  \citenamefont {Rigolin}}]{PhysRevLett.105.095702}%
  \BibitemOpen
  \bibfield  {author} {\bibinfo {author} {\bibfnamefont {T.}~\bibnamefont
  {Werlang}}, \bibinfo {author} {\bibfnamefont {C.}~\bibnamefont {Trippe}},
  \bibinfo {author} {\bibfnamefont {G.~A.~P.}\ \bibnamefont {Ribeiro}}, \ and\
  \bibinfo {author} {\bibfnamefont {G.}~\bibnamefont {Rigolin}},\ }\Doi
  {10.1103/PhysRevLett.105.095702} {\bibfield  {journal} {\bibinfo  {journal}
  {Phys. Rev. Lett.},\ }\textbf {\bibinfo {volume} {105}},\ \bibinfo {pages}
  {095702} (\bibinfo {year} {2010})}\BibitemShut {NoStop}%
\bibitem [{\citenamefont {Maziero}\ \emph {et~al.}(2010)\citenamefont
  {Maziero}, \citenamefont {Guzman}, \citenamefont {C\'eleri}, \citenamefont
  {Sarandy},\ and\ \citenamefont {Serra}}]{PhysRevA.82.012106}%
  \BibitemOpen
  \bibfield  {author} {\bibinfo {author} {\bibfnamefont {J.}~\bibnamefont
  {Maziero}}, \bibinfo {author} {\bibfnamefont {H.~C.}\ \bibnamefont {Guzman}},
  \bibinfo {author} {\bibfnamefont {L.~C.}\ \bibnamefont {C\'eleri}}, \bibinfo
  {author} {\bibfnamefont {M.~S.}\ \bibnamefont {Sarandy}}, \ and\ \bibinfo
  {author} {\bibfnamefont {R.~M.}\ \bibnamefont {Serra}},\ }\Doi
  {10.1103/PhysRevA.82.012106} {\bibfield  {journal} {\bibinfo  {journal}
  {Phys. Rev. A},\ }\textbf {\bibinfo {volume} {82}},\ \bibinfo {pages}
  {012106} (\bibinfo {year} {2010})}\BibitemShut {NoStop}%
\bibitem [{\citenamefont {Werlang}\ \emph {et~al.}(2011)\citenamefont
  {Werlang}, \citenamefont {Ribeiro},\ and\ \citenamefont
  {Rigolin}}]{PhysRevA.83.062334}%
  \BibitemOpen
  \bibfield  {author} {\bibinfo {author} {\bibfnamefont {T.}~\bibnamefont
  {Werlang}}, \bibinfo {author} {\bibfnamefont {G.~A.~P.}\ \bibnamefont
  {Ribeiro}}, \ and\ \bibinfo {author} {\bibfnamefont {G.}~\bibnamefont
  {Rigolin}},\ }\Doi {10.1103/PhysRevA.83.062334} {\bibfield  {journal}
  {\bibinfo  {journal} {Phys. Rev. A},\ }\textbf {\bibinfo {volume} {83}},\
  \bibinfo {pages} {062334} (\bibinfo {year} {2011})}\BibitemShut {NoStop}%
\bibitem [{\citenamefont {Zurek}(2003)}]{PhysRevA.67.012320}%
  \BibitemOpen
  \bibfield  {author} {\bibinfo {author} {\bibfnamefont {W.~H.}\ \bibnamefont
  {Zurek}},\ }\Doi {10.1103/PhysRevA.67.012320} {\bibfield  {journal} {\bibinfo
   {journal} {Phys. Rev. A},\ }\textbf {\bibinfo {volume} {67}},\ \bibinfo
  {pages} {012320} (\bibinfo {year} {2003})}\BibitemShut {NoStop}%
\bibitem [{\citenamefont {Maruyama}\ \emph {et~al.}(2009)\citenamefont
  {Maruyama}, \citenamefont {Nori},\ and\ \citenamefont
  {Vedral}}]{RevModPhys.81.1}%
  \BibitemOpen
  \bibfield  {author} {\bibinfo {author} {\bibfnamefont {K.}~\bibnamefont
  {Maruyama}}, \bibinfo {author} {\bibfnamefont {F.}~\bibnamefont {Nori}}, \
  and\ \bibinfo {author} {\bibfnamefont {V.}~\bibnamefont {Vedral}},\ }\Doi
  {10.1103/RevModPhys.81.1} {\bibfield  {journal} {\bibinfo  {journal} {Rev.
  Mod. Phys.},\ }\textbf {\bibinfo {volume} {81}},\ \bibinfo {pages} {1}
  (\bibinfo {year} {2009})}\BibitemShut {NoStop}%
\bibitem [{\citenamefont {Sarovar}\ \emph {et~al.}(2010)\citenamefont
  {Sarovar}, \citenamefont {Ishizaki}, \citenamefont {Fleming},\ and\
  \citenamefont {Whaley}}]{Sarovar:2010fk}%
  \BibitemOpen
  \bibfield  {author} {\bibinfo {author} {\bibfnamefont {M.}~\bibnamefont
  {Sarovar}}, \bibinfo {author} {\bibfnamefont {A.}~\bibnamefont {Ishizaki}},
  \bibinfo {author} {\bibfnamefont {G.~R.}\ \bibnamefont {Fleming}}, \ and\
  \bibinfo {author} {\bibfnamefont {K.~B.}\ \bibnamefont {Whaley}},\ }\Doi
  {10.1038/NPHYS1652} {\bibfield  {journal} {\bibinfo  {journal} {Nature
  Physics},\ }\textbf {\bibinfo {volume} {6}},\ \bibinfo {pages} {462}
  (\bibinfo {year} {2010})}\BibitemShut {NoStop}%
\bibitem [{\citenamefont {Luo}(2008)}]{PhysRevA.77.042303}%
  \BibitemOpen
  \bibfield  {author} {\bibinfo {author} {\bibfnamefont {S.}~\bibnamefont
  {Luo}},\ }\Doi {10.1103/PhysRevA.77.042303} {\bibfield  {journal} {\bibinfo
  {journal} {Phys. Rev. A},\ }\textbf {\bibinfo {volume} {77}},\ \bibinfo
  {pages} {042303} (\bibinfo {year} {2008})}\BibitemShut {NoStop}%
\bibitem [{\citenamefont {Bennett}\ \emph {et~al.}(1996)\citenamefont
  {Bennett}, \citenamefont {Bernstein}, \citenamefont {Popescu},\ and\
  \citenamefont {Schumacher}}]{PhysRevA.53.2046}%
  \BibitemOpen
  \bibfield  {author} {\bibinfo {author} {\bibfnamefont {C.~H.}\ \bibnamefont
  {Bennett}}, \bibinfo {author} {\bibfnamefont {H.~J.}\ \bibnamefont
  {Bernstein}}, \bibinfo {author} {\bibfnamefont {S.}~\bibnamefont {Popescu}},
  \ and\ \bibinfo {author} {\bibfnamefont {B.}~\bibnamefont {Schumacher}},\
  }\Doi {10.1103/PhysRevA.53.2046} {\bibfield  {journal} {\bibinfo  {journal}
  {Phys. Rev. A},\ }\textbf {\bibinfo {volume} {53}},\ \bibinfo {pages} {2046}
  (\bibinfo {year} {1996})}\BibitemShut {NoStop}%
\bibitem [{\citenamefont {Wootters}(1998)}]{Wootters:1998fk}%
  \BibitemOpen
  \bibfield  {author} {\bibinfo {author} {\bibfnamefont {W.}~\bibnamefont
  {Wootters}},\ }\href@noop {} {\bibfield  {journal} {\bibinfo  {journal}
  {Physical Review Letters},\ }\textbf {\bibinfo {volume} {80}},\ \bibinfo
  {pages} {2245} (\bibinfo {year} {1998})}\BibitemShut {NoStop}%
\bibitem [{\citenamefont {Peng}\ \emph {et~al.}(2005)\citenamefont {Peng},
  \citenamefont {Du},\ and\ \citenamefont {Suter}}]{PhysRevA.71.012307}%
  \BibitemOpen
  \bibfield  {author} {\bibinfo {author} {\bibfnamefont {X.}~\bibnamefont
  {Peng}}, \bibinfo {author} {\bibfnamefont {J.}~\bibnamefont {Du}}, \ and\
  \bibinfo {author} {\bibfnamefont {D.}~\bibnamefont {Suter}},\ }\Doi
  {10.1103/PhysRevA.71.012307} {\bibfield  {journal} {\bibinfo  {journal}
  {Phys. Rev. A},\ }\textbf {\bibinfo {volume} {71}},\ \bibinfo {pages}
  {012307} (\bibinfo {year} {2005})}\BibitemShut {NoStop}%
\bibitem [{\citenamefont {Peng}\ \emph {et~al.}(2010)\citenamefont {Peng},
  \citenamefont {Wu}, \citenamefont {Li}, \citenamefont {Suter},\ and\
  \citenamefont {Du}}]{PhysRevLett.105.240405}%
  \BibitemOpen
  \bibfield  {author} {\bibinfo {author} {\bibfnamefont {X.}~\bibnamefont
  {Peng}}, \bibinfo {author} {\bibfnamefont {S.}~\bibnamefont {Wu}}, \bibinfo
  {author} {\bibfnamefont {J.}~\bibnamefont {Li}}, \bibinfo {author}
  {\bibfnamefont {D.}~\bibnamefont {Suter}}, \ and\ \bibinfo {author}
  {\bibfnamefont {J.}~\bibnamefont {Du}},\ }\Doi
  {10.1103/PhysRevLett.105.240405} {\bibfield  {journal} {\bibinfo  {journal}
  {Phys. Rev. Lett.},\ }\textbf {\bibinfo {volume} {105}},\ \bibinfo {pages}
  {240405} (\bibinfo {year} {2010})}\BibitemShut {NoStop}%
\bibitem [{\citenamefont {Wang}\ \emph {et~al.}(2011)\citenamefont {Wang},
  \citenamefont {Rong}, \citenamefont {Feng}, \citenamefont {Xu}, \citenamefont
  {Chong}, \citenamefont {Su}, \citenamefont {Gong},\ and\ \citenamefont
  {Du}}]{PhysRevLett.106.040501}%
  \BibitemOpen
  \bibfield  {author} {\bibinfo {author} {\bibfnamefont {Y.}~\bibnamefont
  {Wang}}, \bibinfo {author} {\bibfnamefont {X.}~\bibnamefont {Rong}}, \bibinfo
  {author} {\bibfnamefont {P.}~\bibnamefont {Feng}}, \bibinfo {author}
  {\bibfnamefont {W.}~\bibnamefont {Xu}}, \bibinfo {author} {\bibfnamefont
  {B.}~\bibnamefont {Chong}}, \bibinfo {author} {\bibfnamefont {J.-H.}\
  \bibnamefont {Su}}, \bibinfo {author} {\bibfnamefont {J.}~\bibnamefont
  {Gong}}, \ and\ \bibinfo {author} {\bibfnamefont {J.}~\bibnamefont {Du}},\
  }\Doi {10.1103/PhysRevLett.106.040501} {\bibfield  {journal} {\bibinfo
  {journal} {Phys. Rev. Lett.},\ }\textbf {\bibinfo {volume} {106}},\ \bibinfo
  {pages} {040501} (\bibinfo {year} {2011})}\BibitemShut {NoStop}%
\bibitem [{\citenamefont {Tyryshkin}\ \emph {et~al.}(2003)\citenamefont
  {Tyryshkin}, \citenamefont {Lyon}, \citenamefont {Astashkin},\ and\
  \citenamefont {Raitsimring}}]{PhysRevB.68.193207}%
  \BibitemOpen
  \bibfield  {author} {\bibinfo {author} {\bibfnamefont {A.~M.}\ \bibnamefont
  {Tyryshkin}}, \bibinfo {author} {\bibfnamefont {S.~A.}\ \bibnamefont {Lyon}},
  \bibinfo {author} {\bibfnamefont {A.~V.}\ \bibnamefont {Astashkin}}, \ and\
  \bibinfo {author} {\bibfnamefont {A.~M.}\ \bibnamefont {Raitsimring}},\ }\Doi
  {10.1103/PhysRevB.68.193207} {\bibfield  {journal} {\bibinfo  {journal}
  {Phys. Rev. B},\ }\textbf {\bibinfo {volume} {68}},\ \bibinfo {pages}
  {193207} (\bibinfo {year} {2003})}\BibitemShut {NoStop}%
\bibitem [{\citenamefont {Du}\ \emph {et~al.}(2001){\natexlab{b}}\citenamefont
  {Du}, \citenamefont {Shi}, \citenamefont {Wu}, \citenamefont {Zhou},\ and\
  \citenamefont {Han}}]{PhysRevA.63.042302}%
  \BibitemOpen
  \bibfield  {author} {\bibinfo {author} {\bibfnamefont {J.}~\bibnamefont
  {Du}}, \bibinfo {author} {\bibfnamefont {M.}~\bibnamefont {Shi}}, \bibinfo
  {author} {\bibfnamefont {J.}~\bibnamefont {Wu}}, \bibinfo {author}
  {\bibfnamefont {X.}~\bibnamefont {Zhou}}, \ and\ \bibinfo {author}
  {\bibfnamefont {R.}~\bibnamefont {Han}},\ }\Doi {10.1103/PhysRevA.63.042302}
  {\bibfield  {journal} {\bibinfo  {journal} {Phys. Rev. A},\ }\textbf
  {\bibinfo {volume} {63}},\ \bibinfo {pages} {042302} (\bibinfo {year}
  {2001}{\natexlab{b}})}\BibitemShut {NoStop}%
\bibitem [{\citenamefont {Tyryshkin}\ \emph {et~al.}(2006)\citenamefont
  {Tyryshkin}, \citenamefont {Morton}, \citenamefont {Benjamin}, \citenamefont
  {Ardavan}, \citenamefont {Briggs}, \citenamefont {Ager},\ and\ \citenamefont
  {Lyon}}]{Tyryshkin:2006fk}%
  \BibitemOpen
  \bibfield  {author} {\bibinfo {author} {\bibfnamefont {A.}~\bibnamefont
  {Tyryshkin}}, \bibinfo {author} {\bibfnamefont {J.}~\bibnamefont {Morton}},
  \bibinfo {author} {\bibfnamefont {S.}~\bibnamefont {Benjamin}}, \bibinfo
  {author} {\bibfnamefont {A.}~\bibnamefont {Ardavan}}, \bibinfo {author}
  {\bibfnamefont {G.}~\bibnamefont {Briggs}}, \bibinfo {author} {\bibfnamefont
  {J.}~\bibnamefont {Ager}}, \ and\ \bibinfo {author} {\bibfnamefont
  {S.}~\bibnamefont {Lyon}},\ }\Doi {DOI 10.1088/0953-8984/18/21/S06}
  {\bibfield  {journal} {\bibinfo  {journal} {Journal of Physics-Condensed
  Matter},\ }\textbf {\bibinfo {volume} {18}},\ \bibinfo {pages} {S783}
  (\bibinfo {year} {2006})}\BibitemShut {NoStop}%
\bibitem [{\citenamefont {Vandersypen}\ and\ \citenamefont
  {Chuang}(2004)}]{Vandersypen:2004fk}%
  \BibitemOpen
  \bibfield  {author} {\bibinfo {author} {\bibfnamefont {L.}~\bibnamefont
  {Vandersypen}}\ and\ \bibinfo {author} {\bibfnamefont {I.}~\bibnamefont
  {Chuang}},\ }\href@noop {} {\bibfield  {journal} {\bibinfo  {journal}
  {Reviews of Modern Physics},\ }\textbf {\bibinfo {volume} {76}},\ \bibinfo
  {pages} {1037} (\bibinfo {year} {2004})}\BibitemShut {NoStop}%
\bibitem [{\citenamefont {Werlang}\ and\ \citenamefont
  {Rigolin}(2010)}]{PhysRevA.81.044101}%
  \BibitemOpen
  \bibfield  {author} {\bibinfo {author} {\bibfnamefont {T.}~\bibnamefont
  {Werlang}}\ and\ \bibinfo {author} {\bibfnamefont {G.}~\bibnamefont
  {Rigolin}},\ }\Doi {10.1103/PhysRevA.81.044101} {\bibfield  {journal}
  {\bibinfo  {journal} {Phys. Rev. A},\ }\textbf {\bibinfo {volume} {81}},\
  \bibinfo {pages} {044101} (\bibinfo {year} {2010})}\BibitemShut {NoStop}%
\end{thebibliography}%
\end{document}